\documentclass[superscriptaddress,prb,aps,10pt,twocolumn]{revtex4}
\usepackage{graphicx}
\usepackage{graphics}
\usepackage{epsfig}
\usepackage{amsmath}
\usepackage{amssymb}
\usepackage{subfigure}
\usepackage[sort&compress]{natbib} 

\begin{document}
\def \etal{{\it et al}.~}
\def \etaln{{\it et al}.}

\title{Competition between Superconductivity and Weak Localization in Metal-Mixed Ion-Implanted Polymers}

\author{Andrew P. Stephenson}
 \email{aps@physics.uq.edu.au}
 \affiliation{Centre for Organic Photonics and Electronics, School of Mathematics and Physics, University of Queensland, Brisbane QLD 4072, Australia}
\author{Adam P. Micolich}
 \affiliation{School of Physics, University of New South Wales, Sydney NSW 2052, Australia}
\author{Ujjual Divakar}
\affiliation{Centre for Organic Photonics and Electronics, School of Mathematics and Physics, University of Queensland, Brisbane QLD 4072, Australia}
\author{Paul Meredith}
\affiliation{Centre for Organic Photonics and Electronics, School of Mathematics and Physics, University of Queensland, Brisbane QLD 4072, Australia}
\author{Ben J. Powell}
\affiliation{Centre for Organic Photonics and Electronics, School of Mathematics and Physics, University of Queensland, Brisbane QLD 4072, Australia}

\begin{abstract}
We study the effects of varying the pre-implant film thickness and implant temperature on the electrical and superconducting properties of metal-mixed ion-implanted polymers. We show that it is possible to drive a  superconductor-insulator transition in these materials via control of the fabrication parameters. We observe peaks in the magnetoresistance and demonstrate that these are caused by the interplay between superconductivity and weak localization in these films, which occurs due to their granular structure. We compare these magnetoresistance peaks with those seen in unimplanted films and other organic superconductors, and show that they are distinctly different.
\end{abstract}
\date{\today}
\maketitle
\clearpage

\section{Introduction}
For many years, two-dimensional (2D) systems have been the focus of considerable attention due to the interesting interplay between interactions, disorder and dimensionality in determining the electronic ground state.~\cite{Lee1985,Belitz1994} While scaling theory predicted that all 2D systems are driven into an insulating ground state by arbitrarily weak impurity scattering,~\cite{Abrahams1979} it was subsequently shown that superconductivity should survive well into the localized phase.~\cite{Ma1985} From an experimental perspective, superconducting, metallic and insulating ground states have been observed in ultrathin metal films,~\cite{Dynes1978} with transitions between these states induced by tuning the film thickness~\cite{Haviland1989} or applying a magnetic field.~\cite{Hebard1990}

The 2D superconductor-insulator transition has been studied in a variety of ultrathin films with differing compositions (e.g.,
elemental metals, alloys, etc.) and morphologies (e.g., amorphous, crystalline, granular, etc.).~\cite{Goldman1998} Disorder in these films is heavily dependent upon morphology, producing many sample-specific behaviors, for example, quasi-reentrant transitions~\cite{Orr1985,Jaeger1986} and anomalous magnetoresistance peaks.\cite{Hebard1990,Okuma1998,Baturina2004} These behaviors are much more common in granular systems, and thus their observation in a new material system may provide
important clues to its morphology. 

We recently reported a novel superconducting material produced by evaporating a thin film of Sn/Sb alloy onto a polyetheretherketone (PEEK) substrate and subsequently mixing the metal into the PEEK surface using a nitrogen ion-beam.~\cite{Micolich2006} Here we show that  a superconductor-insulator transition can be driven by controlling the thickness of the layer and the implant temperature. We observe peaks in the magnetoresistance of an insulating sample close to this phase transition. We show that these magnetoresistance peaks are caused by the competition between weak localization and superconductivity.

\section{Methods}
The samples studied are produced and measured using methods reported previously.~\cite{Tavenner2004,Micolich2006,Stephenson2009a} To summarize briefly, we commence with cleaned PEEK substrates onto
which a thin film of 19:1 Sn:Sb is deposited by thermal evaporation. For the metal-mixed samples, ion-implantation was then performed using a 0.37~$\mu$Acm$^{-2}$, $50$~keV N$^{+}$ ion-beam that illuminated a circular area $14$~mm in diameter to a dose of $10^{16}$~ions/cm$^2$. During implantation, the sample is mounted on a temperature controlled stage, which is vital to achieving working samples. Electrical contacts were produced by shadow-masked evaporation of $50$~nm Ti followed by $50$~nm Au onto the four corners of each sample, and the sample is cut into a van der Pauw configuration ensuring that the unimplanted regions do not short out measurement of the implanted region. Cu wires are attached to the contacts using In solder. A photograph of a completed sample is shown in the inset to Fig. \ref{fig:NZfig1}(a). Low temperature electrical resistance measurements were carried out using a Keithley 2000 multimeter with the samples mounted in an Oxford Instruments variable temperature insert system capable of temperatures, $T$, between $1.2$ and $200$~K and magnetic fields, $B$, up to $10$~T.

In this paper we report on five samples -- four are metal-mixed and one is not. The four metal-mixed samples form a $2 \times 2$ set with two nominal Sn:Sb alloy thicknesses ($100$~\AA~ and $200$~\AA) and two sample temperatures during implantation ($300$~K and $77$~K). To avoid thickness variations from interfering with studies of implant temperature, the samples for each temperature are cut as pieces from a larger film coated with a specified thickness of Sn:Sb in a single evaporation. The fifth sample is an unimplanted Sn:Sb film with nominal thickness $200$~\AA~ (produced separately from the set of four metal-mixed samples), which provides an interesting counterpoint to the magnetoresistance data obtained from the $100$~\AA~ thick metal-mixed samples.

\section{Results and Discussion}
Before focusing on the key features of the samples, we first make some general comments regarding the sample set that we chose to measure. The electronic properties of metal-mixed polymers can be controlled via a number of parameters involved in the fabrication, including substrate composition; pre-implant metal film thickness and composition; beam energy, current, dose and species; and implantation conditions such as temperature. An exhaustive exploration of this very large, multidimensional parameter space is clearly an onerous task, forcing us to be selective in order to make progress.

For this paper, we have restricted ourselves to a small sample set focused on two key parameters. The first is the pre-implant metal thickness because it provides the easiest control over the conductivity, even though this can be a slightly difficult parameter to control with precision.~\cite{Stephenson2009a} The second is the implant temperature, which we believe provides some control over the disorder of the resulting film, as we will show in the next section. A more extensive study of the role of the fabrication and ion-implantation parameters in determining the sample conductivity will be the subject of a separate, forthcoming paper.

\subsection{The effect of implantation temperature}

\begin{figure*}
	\includegraphics[width=0.99\textwidth]{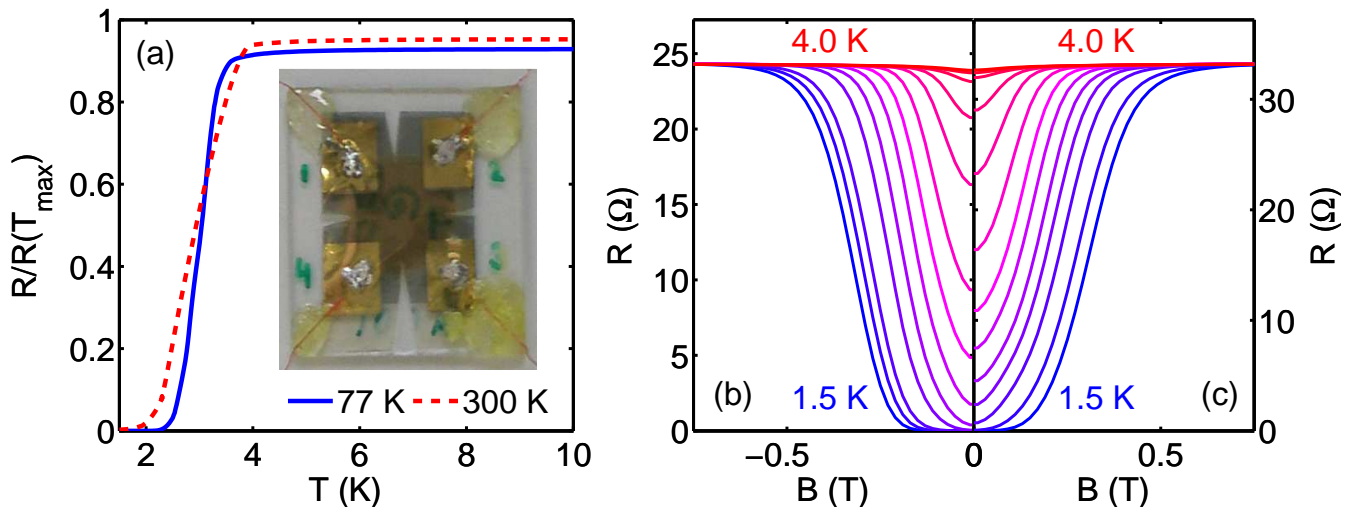}
		\caption{(Color online) (a) The normalized four-terminal resistance $R(T)/R(T_\textrm{max})$ versus temperature $T$ for $200$~\AA~ Sn:Sb films implanted at $77$~K (solid blue line) and $300$~K (dashed red line). The critical temperature $T_{c}$ and transition width $\Delta T$ are $3.0$~K and $0.63$~K for the $77$~K sample, and $2.9$~K and $1.0$~K for the $300$~K sample. The higher $R(T)$ for $T > T_{c}$, reduced $T_{c}$ and larger $\Delta T$ point to a higher disorder for the $300$~K sample. (Inset) A photograph of a typical ion-implanted sample. Panels (b) and (c) show the resistance $R$ versus applied perpendicular magnetic field $B$ at temperatures $T$ ranging between $1.5$ and $4.0$~K for the $77$~K and $300$~K samples respectively. At $T = 1.5$~K, the critical field $B_{c}$ and transition width $\Delta B$ are $0.33$~T and $0.19$~T for the $77$~K sample, and $0.31$~T and $0.24$~T for the $300$~K sample. The lower $B_{c}$ and larger $\Delta B$ again confirm the higher disorder in the $300$~K sample.} 
	\label{fig:NZfig1}
\end{figure*}

We will start by considering the two $200$~\AA~ metal-mixed samples, which exhibit a metallic temperature dependence for temperatures greater than the critical temperature $T_{c}$ and a clean transition to a global (i.e., sample-wide) zero resistance state. Comparing the resistance measured between the four contact pairs along the sides of the sample and two pairs running diagonally, both of these samples are relatively isotropic (cf. $100$~\AA~ samples discussed in Section IIIB). In Fig. \ref{fig:NZfig1}(a) we present the normalized resistance $R(T)/R(T_\textrm{max})$, where $T_\textrm{max} = 202.6$~K, measured in a four-terminal configuration for the $200$~\AA~ samples implanted at $77$~K (solid blue line) and $300$~K (dashed red line). The resistance at $T_\textrm{max}$ is $24.2~\Omega$ for the $77$~K sample and $33.1~\Omega$ for the $300$~K sample, which also has the greater normalized resistance for $T > T_{c}$. Additionally, the $300$~K sample has the lower $T_{c}$ and larger transition width, almost double that of the $77$~K sample, as expected for a sample with a higher normal resistance and higher disorder.~\cite{Strongin1970,Raffy1983,Graybeal1984} Further evidence for the relationship between disorder and implant temperature is provided by the magnetic field data presented in panels (b) and (c) of Fig.~\ref{fig:NZfig1} for samples implanted at $77$~K and $300$~K respectively. Considering, for example, the data at $T=1.5$~K, the critical field, $B_{c}$, is lower and the transition width, $\Delta B$, is larger for the $300$~K sample, again pointing to higher disorder in this sample. This dependence of the sample properties on implant temperature points to an ability to fine-tune the sample properties via the implant parameters, over and above the tuning provided by the metal thickness. This provides an incredible versatility to these metal-mixed polymers as an electronic materials system, as we will demonstrate systematically in a forthcoming publication.

\begin{figure}
\centering
\includegraphics[width=0.48\textwidth]{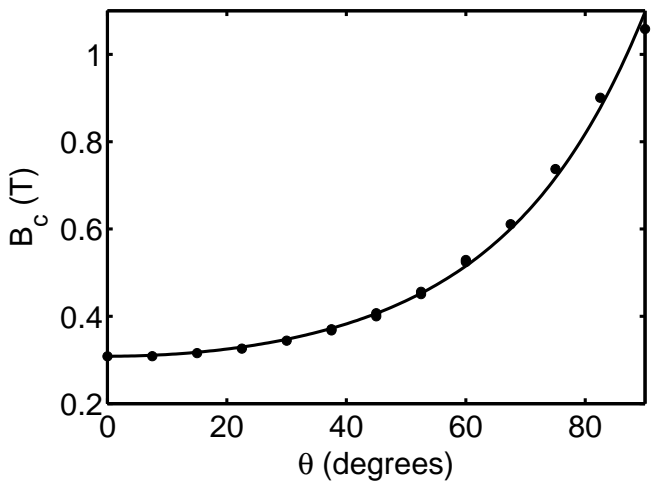}
\caption{The critical field, $B_{c}$, versus the angle, $\theta$, of the field relative to the film normal for the 200~\AA~ sample implanted at 77~K. The solid line is a fit of Eqn.~\ref{2Dsuperconductor} to the experimental data, and the quality of this fit demonstrates that this sample is two-dimensional.} 
\label{BcvTheta}
\end{figure}

Focusing on the 200~\AA~ sample deposited at 77~K [Fig.~\ref{fig:NZfig1}(b)], we have measured the angular dependence of the critical field to determine its dimensionality. For a two-dimensional superconductor, the angular dependence should have the following form:\cite{Tinkham1963}
\begin{equation}
\bigg \lvert \frac{B_{c}(\theta) \sin\theta}{B_{c}^{\perp}} \bigg \rvert + \left( \frac{B_{c}(\theta) \cos\theta}{B_{c}^{\|}} \right)^2 = 1,
\label{2Dsuperconductor}
\end{equation}where $\theta$ is the angle of the magnetic field relative to the film, and $B^{\perp}_{c}$ and $B^{\parallel}_{c}$ are the critical fields obtained when the magnetic field is normal to the film ($\theta = 90^{\circ}$) and in the plane of the film ($\theta = 0^{\circ}$), respectively. In Fig.~\ref{BcvTheta} we show the measured critical field $B_{c}$ versus angle $\theta$, with the solid line presenting a fit of Eqn.~\ref{2Dsuperconductor} to the data. We obtain $B_{c}$ as the field at which the sample resistance is half of that obtained in the normal state. The excellent fit to the data provided by Eqn.~\ref{2Dsuperconductor} indicates that our sample is two-dimensional, and since this is the thickest and cleanest of the metal-mixed samples that we study, this implies that our other samples are also in the 2D limit.

\subsection{Crossing over to the insulating side}

The resistivity of the $100$~\AA~ samples is much higher, commensurate with their reduced metal thickness.~\cite{Dynes1978}
Both of these samples are in the insulating regime (i.e., resistivity increases with decreasing $T$), however an unfortunate
side-effect is that the electronic properties of these samples are significantly more anisotropic. This makes it impossible to sensibly obtain the resistivity through four-terminal measurements, and hence all resistance measurements that we report for the $100$~\AA~ samples are two-terminal measurements. Due to the strong anisotropy, the effect of implant temperature on the resistance is not quite as obvious in these samples. The corner to corner room-temperature resistances vary from $\sim 22$ to $135$~k$\Omega$ for the $300$~K sample and from $\sim 13$ to $900$~k$\Omega$ in the corresponding $77$~K sample. The lowest resistance is measured in the $77$~K sample, and is lower by a factor of $\sim 2$ than the lowest resistance in the $300$~K sample. We will now focus on the $100$~\AA~ sample implanted at $300$~K, since the higher disorder in this sample strengthens the effects that we will discuss.

\begin{figure}
\centering
\includegraphics[width=0.47\textwidth]{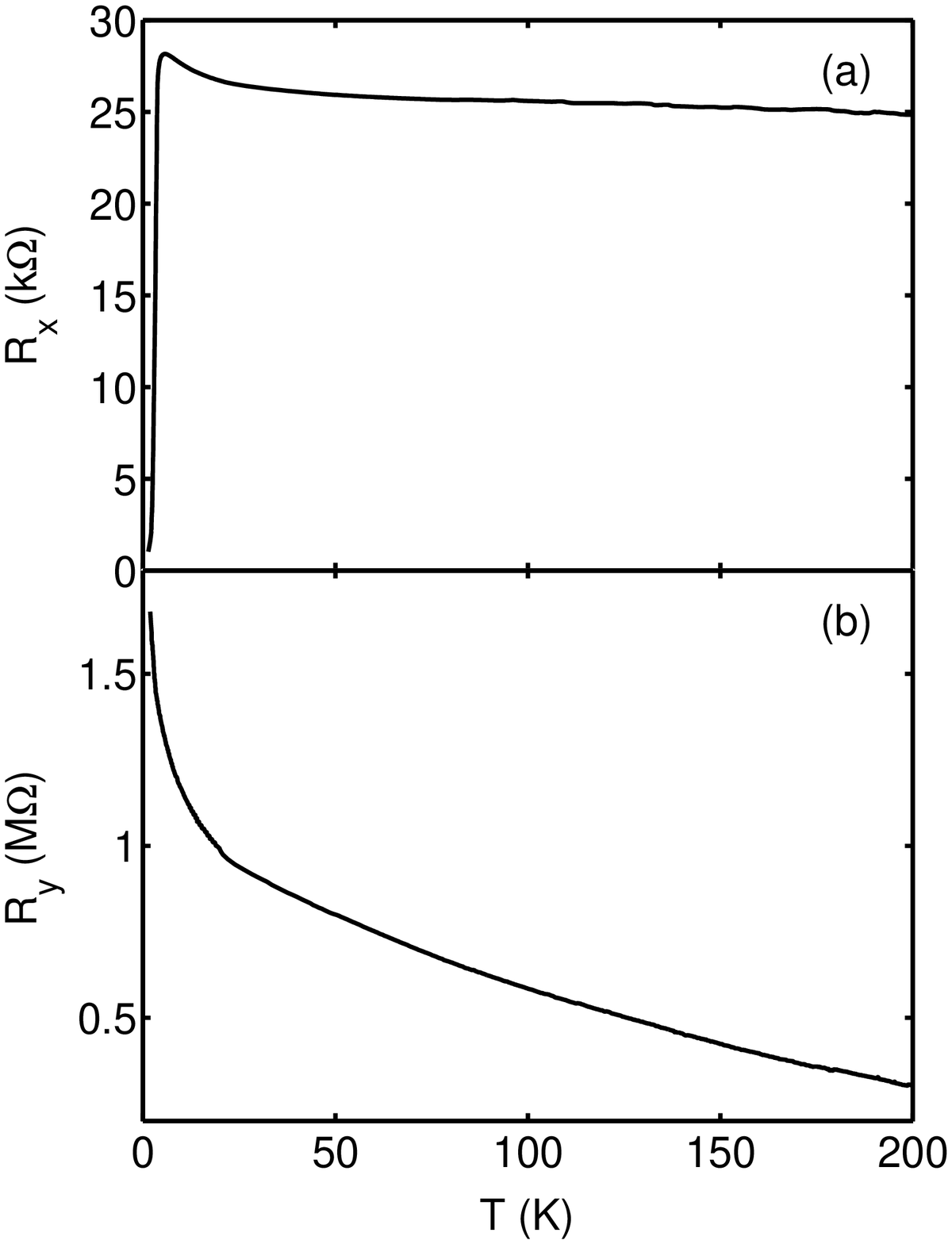}
\caption{The two-terminal resistance $R$ versus $T$ measured (a) along the $x$-direction, and (b) along the $y$-direction for the $100$~\AA~ sample implanted at $300$~K. These two measurements along perpendicular edges of the sample utilize a common contact.}
\label{fig:NZfig2}
\end{figure}

In Fig.~\ref{fig:NZfig2} we present two-terminal resistance measurements for two perpendicular edges of the 300~K. In the following we denote the lower resistance direction as the $x$-direction, $R_{x}$, [see Fig.~\ref{fig:NZfig2}(a)] and the higher resistance direction as the $y$-direction, $R_{y}$ [see Fig.~\ref{fig:NZfig2}(b)]. Considering Fig.~\ref{fig:NZfig2}(a) first, the sample is clearly insulating along the $x$-direction (increasing $R$ with decreasing $T$ for $T > T_{c}$), but undergoes an incomplete superconducting transition at a temperature of approximately $3.2$~K. A sample-wide zero-resistance state could not be reached within the temperature range available with our cryostat [$R(T = 1.6$~K$)\sim 100~\Omega$], and it is unclear whether one could be attained by going to lower temperatures. Such incomplete superconducting transitions are common in granular metal films on the insulating side near to the metal-insulator transition.~\cite{Kobayashi1981, Orr1985, White1986, Kunchur1987, Jaeger1989} Similar quasi-reentrant transitions have also been observed in granular cuprate samples~\cite{Gerber1990, Gray1993} and organic superconductors.~\cite{Kartsovnik1999}

In contrast, along the $y$-direction in this sample [see Fig. \ref{fig:NZfig2}(b)] there is no superconducting transition down to $T = 1.6$~K. The resistance starts $\sim 16$ times higher than that in the $x$-direction at $T = 200$~K and continues to increase as $T$ is reduced, reaching $1.7$~M$\Omega$ at $T = 1.6$~K. Such a strong anisotropy is not uncommon in metal-mixed samples in the insulating regime. The observed anisotropy in this material can be explained with a granular model where some grains are insulating, while others are superconducting and may be coupled via the Josephson or proximity effects. Anisotropies in the grain distribution result in there being no percolation path for superconductivity in the $y$-direction, whereas in the $x$-direction a percolation path does exist or is very weakly broken [consistent with the small, but non-zero, resistance in this direction, Fig. \ref{fig:NZfig2}(a)]. A natural prediction of such a model is that some signatures of the superconducting grains should remain in the measured resistance along the $y$-direction, and these signatures are observed, as we will demonstrate below. 

\begin{figure*}
\centering
\includegraphics[width=0.48\textwidth]{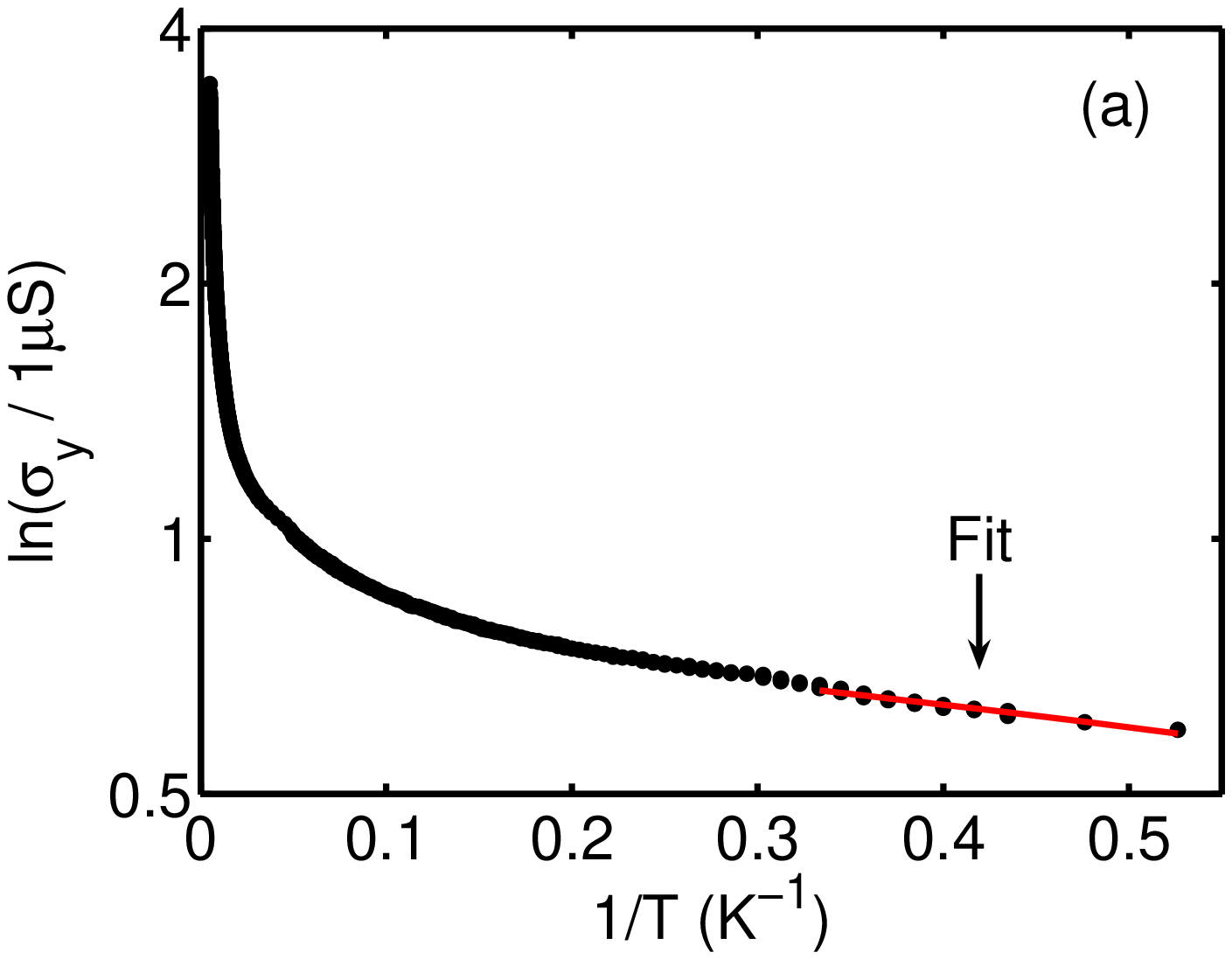}
\includegraphics[width=0.48\textwidth]{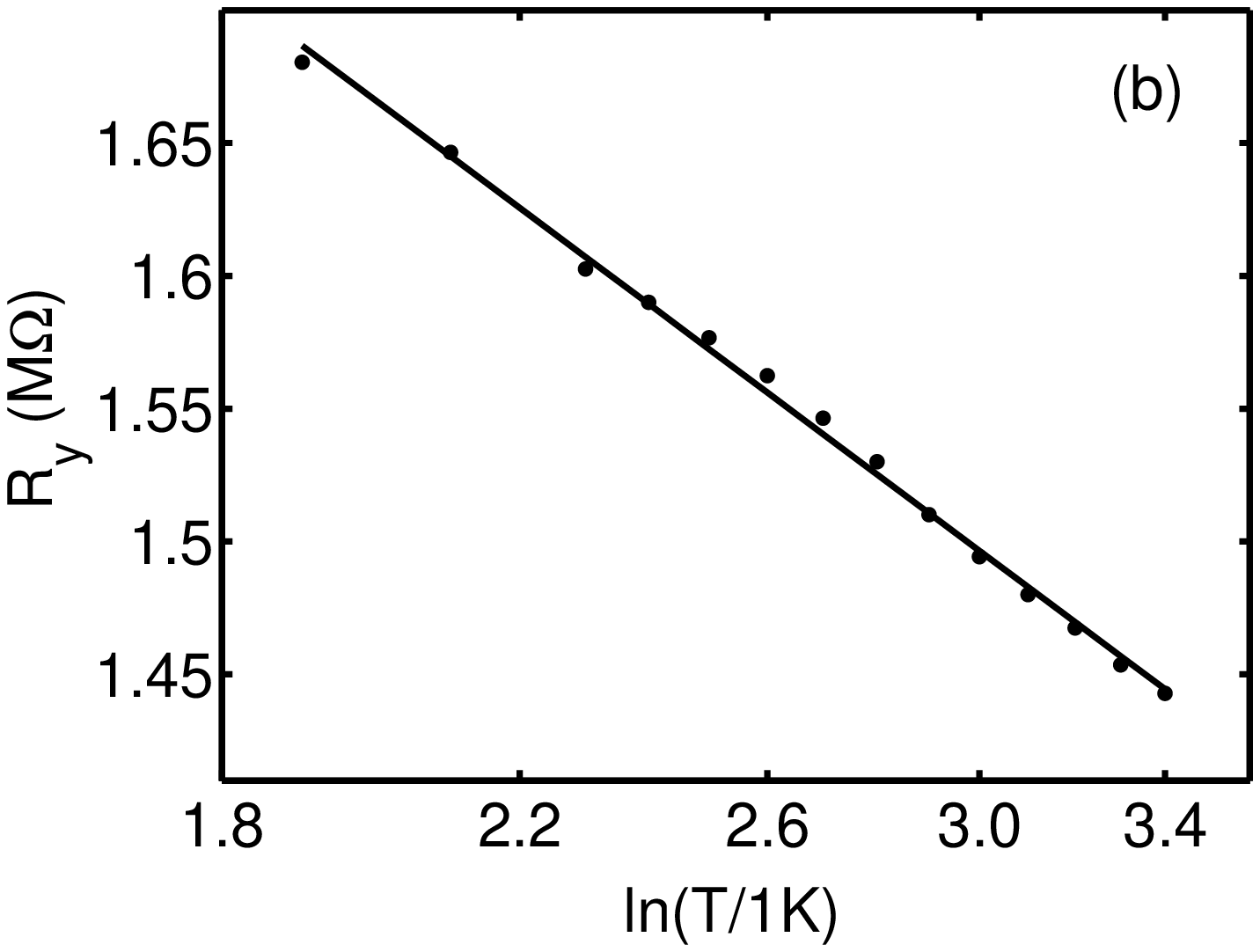}
\caption{(Color online) (a) An Arrhenius plot of $\ln\sigma_{y}$ versus $1/T$, where $\sigma_{y} = R_{y}^{-1}$, and (b) a plot of $R_{y}$ versus $\ln T$, for the data presented in Fig.~\ref{fig:NZfig2}(b). An Arrhenius model only fits the data in (a) for $T < 4$~K and gives a value for the energy gap $\Delta /k_{B} \leq 1~$K. In contrast, the linear dependence in (b) suggests that the origin of the insulating behavior in this sample is due to weak localization.} 
\label{arrheniusWL}
\end{figure*}

To understand the origin of this insulating behavior, in Fig. \ref{arrheniusWL} we fit the data in Fig. \ref{fig:NZfig2}(b) to two models. Firstly, in Fig. \ref{arrheniusWL}(a) we plot the data in Fig. \ref{fig:NZfig2}(b) on a graph of $\ln\sigma_{y}$ versus $1/T$ and attempt to fit an Arrhenius model to the data [i.e., $R \propto$ exp$(-\Delta~/~k_{B}T)$ where $\Delta$ is an insulating gap]. As Fig. \ref{arrheniusWL}(a) shows, this model only fits well for $T < 4$~K. However, this fit gives a value for the gap $\Delta / k_{B} \leq 1~$K, indicating that an opening of an energy gap at the Fermi level in our system cannot be responsible for the insulating behavior. As a second alternative, we consider the possibility that the insulating behavior is instead due to weak localization.\cite{Bergmann1984} In a 2D system weak localization leads to a resistance that is proportional to $\ln T$. In a 2D system, weak localization leads to a resistance that is proportional to $\ln T$. The angular dependence study for our thickest sample shows clear 2D behavior, and so we expect this logarithmic temperature dependence to appear in any of our samples in which weak localization also appears. For example, in Fig.~\ref{arrheniusWL}(b) we plot $R_{y}$ versus $\ln T$ for the 100~\AA~ sample deposited at 300~K, and a clear linear trend is observed consistent with weak localization.

\begin{figure}
\centering
\includegraphics[width=0.48\textwidth]{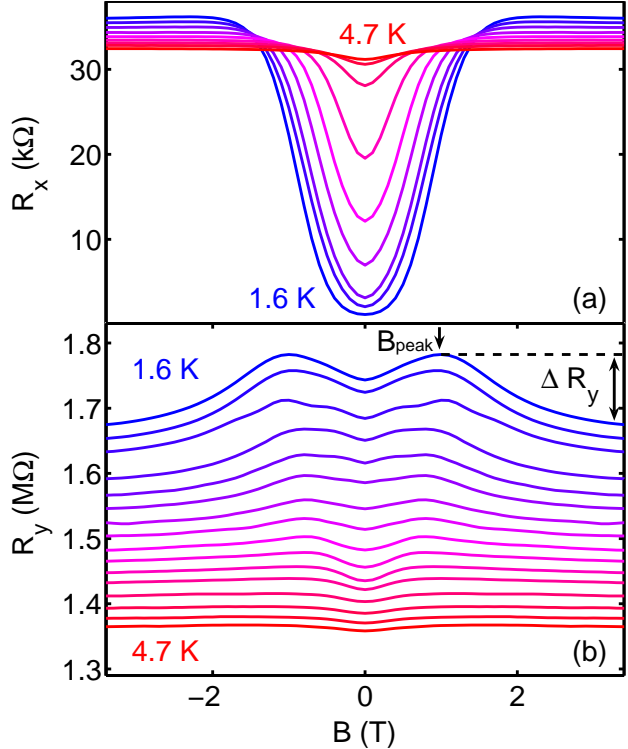}
\caption{(Color online) (a) $R_{x}$ and (b) $R_{y}$ as a function of applied field, $B$, at several different temperatures for the $100$~\AA~ sample implanted at $300$~K. The $R_{x}$ data has a deep minima centered on $B = 0$ that does not reach zero, indicating that the superconductivity in this sample is local and not global. In contrast, the $R_{y}$ data shows a broad negative magnetoresistance (peak) that diminishes with temperature, consistent with weak localization. The superimposed positive magnetoresistance feature (minima) is due to local superconductivity in the sample, and has the same width as the minima in (a). $B_\textrm{peak}$ and $\Delta R_y$ for $T=1.6$~K are indicated in (b).} 
\label{fig:NZfig4}
\end{figure}

Weak localization in 2D systems is also characterised by a negative magnetoresistance (i.e., a resistance peak at $B = 0$).~\cite{Bergmann1984} In Fig.~\ref{fig:NZfig4}(a) and (b) we plot the magnetoresistance $R_{x}(B)$ and $R_{y}(B)$, respectively, at a range of temperatures for the $100$~\AA~ sample implanted at $300$~K. Concomitant with the temperature dependence of $R_{x}$ presented in Fig. \ref{fig:NZfig2}(a), the $R_{x}(B)$ data in Fig.~\ref{fig:NZfig4}(a) features a deep minimum centered at $B = 0$ and a field-induced transition to a normal state at a critical field $B_{c} = 0.9$~T. This transition is relatively wide ($\Delta B_{c} \sim 0.8$~T) at $T = 1.6$~K, and the minimum rises rapidly as the temperature is increased. In each case, however, the resistance becomes field-independent for $|B| \gtrsim 1.5$~T indicating the complete quenching of superconductivity in the sample. The magnetoresistance data presented in Fig.~\ref{fig:NZfig4}(a) is quite similar to that observed in our other superconducting films [e.g., the $200$~\AA~ sample in Figs.~\ref{fig:NZfig1}(b, c)] except that in those samples zero resistance is achieved. The absence of a zero resistance state in Fig. \ref{fig:NZfig4}(a) indicates that a sample-wide superconducting state has not been attained despite clear evidence of local superconductivity.

In contrast, the magnetoresistance along the $y$-direction [see Fig.~\ref{fig:NZfig4}(b)] shows the typical characteristic of weak localization -- a broad peak in the resistance centered at $B = 0$ with a characteristic half-width of order 3~T. The magnitude of this negative magnetoresistance diminishes with increasing temperature, as expected given that weak localization is a quantum interference phenomenon. Positive magnetoresistance is also observed at smaller field scales, and we attribute this to local superconductivity in the sample. The crossover from positive to negative magnetoresistance that occurs at $B \sim 1$~T in Fig. \ref{fig:NZfig4}(b), coincides with the field-induced suppression of superconductivity shown in Fig. \ref{fig:NZfig4}(a), adding support for this explanation for the $B = 0$ minimum in $R_{y}$. The behavior of the resulting magnetoresistance peaks is quite interesting. The field at which the peak magnetoresistance is observed, $B_\textrm{peak}$, is only weakly dependent on temperature, and may be non-monotonic, however, it is difficult to make this statement definitively due to peak broadening as the temperature is elevated. 

\begin{figure}
\centering
\includegraphics[width=0.48\textwidth]{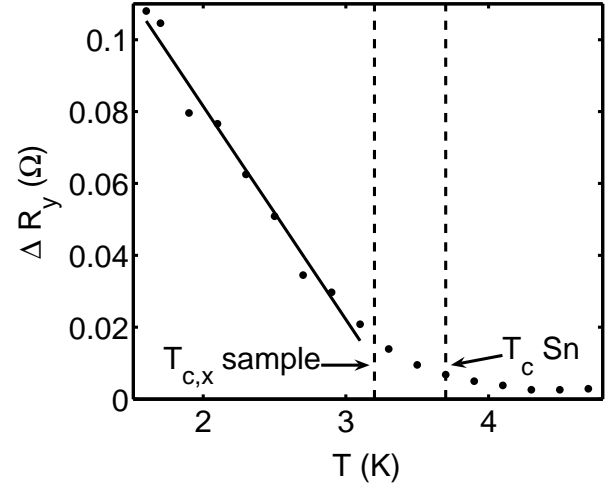}
\caption{The peak resistance, $\Delta R_y=R_y(T,B_\textrm{peak})-R_y(T,B_\textrm{max})$, for data in Fig.~\ref{fig:NZfig4}(b) (the high resistance direction of the 100~\AA~ sample implanted at room temperature) as a function of temperature.  ($B_\textrm{max}=3.5$~T is the maximum field studied.) The behavior of the peaks does not change significantly at the temperature where the resistive transition is observed in the $x$-direction (marked in the figure as `$T_{c,x}$ sample') and the peaks continue up until at least the $T_c$ of bulk tin, also marked in the figure, consistent with a granular structure. } 
\label{fig:7}
\end{figure}

Defining the peak's field location is straightforward but quantifying its height requires a little more consideration. The resistance becomes constant in $B$ at sufficiently high fields as the effects of superconductivity and weak localization are quenched. Hence it makes more sense to reference the peak height to the resistance at the maximum measured field $R(B_\textrm{max})$, than to $R(B = 0)$, for example. This is particularly clear in Fig.~\ref{fig:5}, where we use the same definition to quantify the peak height. Thus we define the peak height $\Delta R=R(T,B_\textrm{peak})-R(T,B_\textrm{max})$. In Fig.~\ref{fig:7} we show the temperature dependence of $\Delta R_y$, the peak resistance obtained in the $y$-direction data in Fig.~\ref{fig:NZfig4}(b). No change in the magnetoresistance peaks is observed at the resistive critical temperature in the $x$-direction, and the peaks are observed at least up to the critical temperature of bulk tin. This is consistent with a granular structure in which different grains become superconducting at slightly different temperatures, beginning at about the $T_c$ for bulk tin. The competition between superconductivity and weak localization in this sample is indicative of a highly disordered and very anisotropic granular metallic film. We attribute the severity of the electrical inhomogeneity to the system's close proximity to a sharp metal-insulator transition. The precise nature of the coupling between the grains is unclear and will require further work. That said, we expect the coupling to be dependent on the nature of the carbonized polymer matrix created by the ion-beam,\cite{Osaheni1992} which fills the space between the grains, as well as the size-distribution and morphology of the grains themselves. However, on the weight of evidence presented here and in our earlier work,\cite{Micolich2006,Stephenson2009a} it is clear that this material system is granular.

\subsection{Weak localization in unimplanted films with metallic conductivity}
We conclude by considering some data from an unimplanted sample with a much higher conductivity, which provides an interesting
counterpoint to the data presented for the implanted samples. The nominal thickness of this sample is $200$~\AA. It was evaporated in a separate batch to the $200$~\AA~ implanted samples and has a lower corner-to-corner resistance. While at first sight this might be attributed to this sample being thicker, it should be remembered that the implantation process spreads the evaporated film by up to ten times its original thickness into the PEEK substrate. This leads to some loss of metal due to sputtering,~\cite{Tavenner2004, Micolich2006} which is the primary cause of the increased resistance after implantation. The low and isotropic resistance in this sample makes it ideal for four-terminal measurements.

\begin{figure}
\centering
\includegraphics[width=0.48\textwidth]{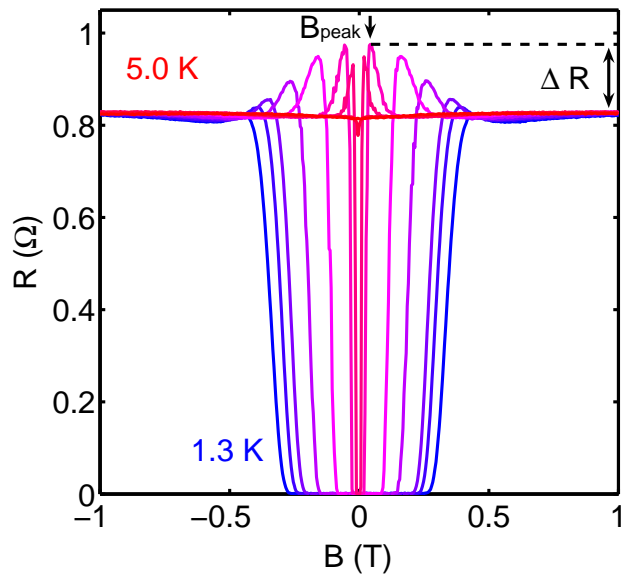}
\caption{(Color online) Four terminal magnetoresistance $R$ at various temperatures between $1.3$ and $5.0$~K for the unimplanted $200$~\AA~ sample. Of particular note are the `peaks' in the magnetoresistance on the normal side of the field-induced superconducting-normal transition (cf. Figs.~1 and 2 in Ref~\onlinecite{Zuo1996}). $B_\textrm{peak}$ and $\Delta R$ for $T=3.5$~K are indicated in (b)} 
\label{fig:5}
\end{figure}

Fig. \ref{fig:5} shows the measured four-terminal magnetoresistance for this sample at a variety of temperatures between $1.3$ and $5.0$~K. Despite having a resistance that is six orders of magnitude smaller than that reported in Fig.~\ref{fig:NZfig4}(b), a negative magnetoresistance is still observed. The natural reaction is that this is also weak localization, since the appearance of weak localization in low resistance thin films is certainly not unusual.~\cite{Bergmann1984} The central minima that we observed in Fig.~\ref{fig:5} due to superconductivity appears as a broad, flat-bottomed minima with zero resistance, very similar to that in Fig. \ref{fig:NZfig1}(b), indicating an electrically continuous, global superconducting state in this sample. This is not surprising given this sample's much lower normal resistance. Combining these negative and positive magnetoresistance contributions together results in the appearance of `peaks' in the magnetoresistance at the point of the field-induced superconductor-normal transition. However, it is not quite so straightforward to attribute these peaks to competition between weak localization and superconductivity, because the question needs to be asked why similar peaks do not occur in the implanted samples [see Figs.~\ref{fig:NZfig1}(b), ~\ref{fig:NZfig1}(c) and \ref{fig:NZfig4}(a)]?

\begin{figure}
\centering
\includegraphics[width=0.48\textwidth]{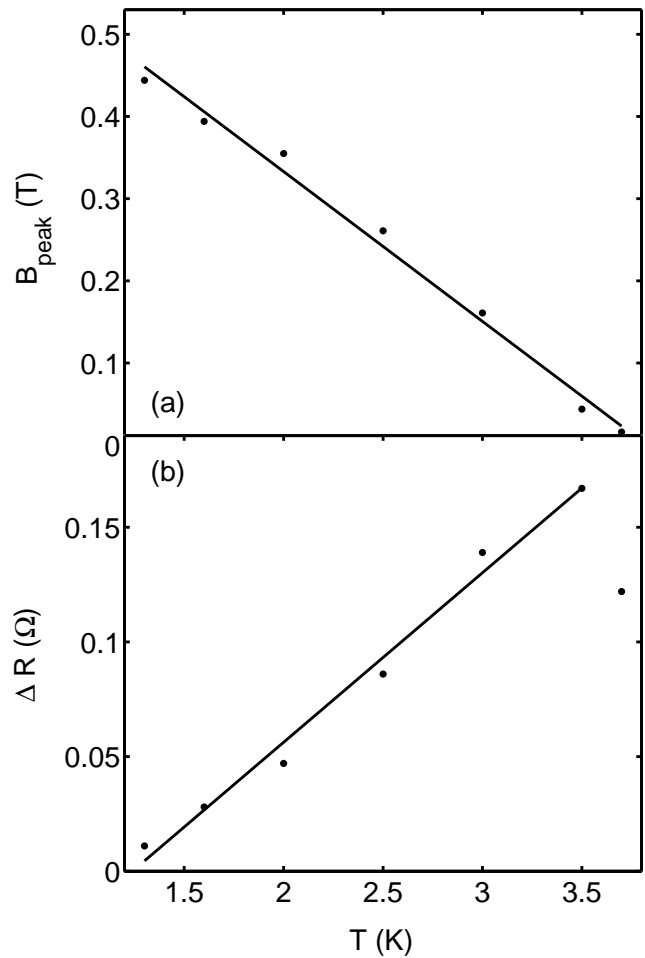}
\caption{(a) Location of the peaks in magnetic field, $B_\textrm{peak}$, and (b) the peak resistance defined as $\Delta R = R(T,B_\textrm{peak}) - R(T,B_\textrm{max})$, where $B_\textrm{max}=1.0$~T, for data in Fig.~\ref{fig:5} as a function of temperature, for comparison with data in Refs.~\onlinecite{Zuo1996} and ~\onlinecite{Zuo1999}.} 
\label{fig:6}
\end{figure}

A simple argument would be that the implantation spreads the film into the substrate, increasing its thickness and making it
three-dimensional. However, this leads to significant chemical binding between the metallic species and the polymer,~\cite{Tavenner2004,Micolich2006} which should reduce the free electron density, increasing the Fermi wavelength and maintaining the 2D limit. Further, this is inconstant with the angular dependence of the critical field shown in Fig.~\ref{BcvTheta}. An interesting alternative to consider is that the peaks in the unimplanted sample are not caused by weak localization at all. Remarkably similar magnetoresistance peaks are observed in data obtained by Zuo {\it et al.} for the quasi-2D organic superconductor $\kappa$-(BEDT-TTF)$_{2}$Cu(NCS)$_{2}$.~\cite{Zuo1996, Zuo1999} 

The parallels between these two effects go beyond the similarities that are obvious to the naked eye. The field at which the peak occurs, $B_\textrm{peak}$,  decreases linearly with temperature as shown in Fig. \ref{fig:6}(a).  Further, the peak resistance, $\Delta R=R(T,B_\textrm{peak})-R(T,B_\textrm{max})$ where $B_\textrm{max}=1.0$~T, increases with increasing temperature as shown in Fig.~\ref{fig:6}(b). Zuo \etal reported both of these effects in $\kappa$-(BEDT-TTF)$_{2}$Cu(NCS)$_{2}$, cf. Fig. 3 of Ref.~\onlinecite{Zuo1996} and Fig. 2 (inset) of Ref.~\onlinecite{Zuo1999}. The differing sign of the gradients for the data in Figs.~\ref{fig:7} and \ref{fig:6}(b) is consistent with the electrical properties (insulating versus metallic) of these two samples. Further, the magnetoresistance peaks in the unimplanted sample [Fig.~\ref{fig:6}(b)] are only observed below the superconducting critical temperature of bulk tin, suggesting that the magnetoresistance peaks are intimately connected with the superconductivity. Zuo {\it et al.} attributed the magnetoresistance peaks in $\kappa$-(BEDT-TTF)$_{2}$Cu(NCS)$_{2}$ to lattice distortion by strong coupling to fluctuating vortices,~\cite{Zuo1996} however, other mechanisms involving dissipation and Josephson-junction effects have also been suggested.\cite{Singleton2000} Further, extensive studies of the role of disorder in these materials have not shown any other signs of weak localization.\cite{Analytis2006,Powell2004} 

Given the very different behavior of the magnetoresistance peaks in the implanted and unimplanted films, it seems reasonable to suggest that different physics may well be at play. It is perhaps dangerous to suppose that the magnetoresistance peaks in our unimplanted films have the same origin as that in $\kappa$-(BEDT-TTF)$_{2}$Cu(NCS)$_{2}$ without much more solid physical evidence, given the important physical differences between the two materials systems.\cite{Micolich2006,Stephenson2009a,Powell2006} However, the commonalities in the data are tantalizing, and further studies of this phenomenon in both systems are certainly called for.

\section{Conclusion}

We have studied the magnetoresistance of  Sn-Sb metal-mixed polymers close to the superconductor-insulator transition. We have shown that close to this transition these materials are highly anisotropic, consistent with a granular model of their structure. There is clear evidence for weak localization in both the temperature dependence of the resistivity and the magnetoresistance. However, weak localization competes with superconductivity, leading to peaks in the magnetoresistance. These magnetoresistance peaks differ in a number of important ways from the peaks we have observed in the magnetoresistance of unimplanted films of Sn-Sb on plastic substrates. It is not yet clear whether this is because fundamentally different physics is at play, or simply because the unimplanted films are much better metals. Intriguingly there are strong similarities between the magnetoresistance of the unimplanted films and that of $\kappa$-(BEDT-TTF)$_{2}$Cu(NCS)$_{2}$, which is a bulk layered crystal.\cite{Zuo1996,Zuo1999, Powell2006}

\section{Acknowledgments}
We would like to thank the Crown Research Organisation for ion-implanting our samples. This work was supported by the
Australian Research Council (ARC) under the Discovery scheme (project number DP0559215). APS received a Short Term Travel Grant
from the Australian Research Council Nanotechnology Network. APM thanks the Division of Solid State Physics at Lund University, Sweden for hospitality during the writing of this paper. PM acknowledges the Queensland Government for the funding through the Smart State Senior Fellowship Program. BJP is supported by the ARC under the Queen Elizabeth II program (project number DP0878523).

\bibliographystyle{apsrev}

\begin{thebibliography}{34}
\expandafter\ifx\csname natexlab\endcsname\relax\def\natexlab#1{#1}\fi
\expandafter\ifx\csname bibnamefont\endcsname\relax
  \def\bibnamefont#1{#1}\fi
\expandafter\ifx\csname bibfnamefont\endcsname\relax
  \def\bibfnamefont#1{#1}\fi
\expandafter\ifx\csname citenamefont\endcsname\relax
  \def\citenamefont#1{#1}\fi
\expandafter\ifx\csname url\endcsname\relax
  \def\url#1{\texttt{#1}}\fi
\expandafter\ifx\csname urlprefix\endcsname\relax\def\urlprefix{URL }\fi
\providecommand{\bibinfo}[2]{#2}
\providecommand{\eprint}[2][]{\url{#2}}

\bibitem[{\citenamefont{Lee and Ramakrishnan}(1985)}]{Lee1985}
\bibinfo{author}{\bibfnamefont{P.~A.} \bibnamefont{Lee}} \bibnamefont{and}
  \bibinfo{author}{\bibfnamefont{T.~V.} \bibnamefont{Ramakrishnan}},
  \bibinfo{journal}{Rev. Mod. Phys.} \textbf{\bibinfo{volume}{57}},
  \bibinfo{pages}{287} (\bibinfo{year}{1985}).

\bibitem[{\citenamefont{Belitz and Kirkpatrick}(1994)}]{Belitz1994}
\bibinfo{author}{\bibfnamefont{D.}~\bibnamefont{Belitz}} \bibnamefont{and}
  \bibinfo{author}{\bibfnamefont{T.~R.} \bibnamefont{Kirkpatrick}},
  \bibinfo{journal}{Rev. Mod. Phys.} \textbf{\bibinfo{volume}{66}},
  \bibinfo{pages}{261} (\bibinfo{year}{1994}).

\bibitem[{\citenamefont{Abrahams et~al.}(1979)\citenamefont{Abrahams, Anderson,
  Licciardello, and Ramakrishnan}}]{Abrahams1979}
\bibinfo{author}{\bibfnamefont{E.}~\bibnamefont{Abrahams}},
  \bibinfo{author}{\bibfnamefont{P.~W.} \bibnamefont{Anderson}},
  \bibinfo{author}{\bibfnamefont{D.~C.} \bibnamefont{Licciardello}},
  \bibnamefont{and} \bibinfo{author}{\bibfnamefont{T.~V.}
  \bibnamefont{Ramakrishnan}}, \bibinfo{journal}{Phys. Rev. Lett.}
  \textbf{\bibinfo{volume}{42}}, \bibinfo{pages}{673} (\bibinfo{year}{1979}).

\bibitem[{\citenamefont{Ma and Lee}(1985)}]{Ma1985}
\bibinfo{author}{\bibfnamefont{M.}~\bibnamefont{Ma}} \bibnamefont{and}
  \bibinfo{author}{\bibfnamefont{P.~A.} \bibnamefont{Lee}},
  \bibinfo{journal}{Phys. Rev. B} \textbf{\bibinfo{volume}{32}},
  \bibinfo{pages}{5658} (\bibinfo{year}{1985}).

\bibitem[{\citenamefont{Dynes et~al.}(1978)\citenamefont{Dynes, Garno, and
  Rowell}}]{Dynes1978}
\bibinfo{author}{\bibfnamefont{R.~C.} \bibnamefont{Dynes}},
  \bibinfo{author}{\bibfnamefont{J.~P.} \bibnamefont{Garno}}, \bibnamefont{and}
  \bibinfo{author}{\bibfnamefont{J.~M.} \bibnamefont{Rowell}},
  \bibinfo{journal}{Phys. Rev. Lett.} \textbf{\bibinfo{volume}{40}},
  \bibinfo{pages}{479} (\bibinfo{year}{1978}).

\bibitem[{\citenamefont{Haviland et~al.}(1989)\citenamefont{Haviland, Liu, and
  Goldman}}]{Haviland1989}
\bibinfo{author}{\bibfnamefont{D.~B.} \bibnamefont{Haviland}},
  \bibinfo{author}{\bibfnamefont{Y.}~\bibnamefont{Liu}}, \bibnamefont{and}
  \bibinfo{author}{\bibfnamefont{A.~M.} \bibnamefont{Goldman}},
  \bibinfo{journal}{Phys. Rev. Lett.} \textbf{\bibinfo{volume}{62}},
  \bibinfo{pages}{2180} (\bibinfo{year}{1989}).

\bibitem[{\citenamefont{Hebard and Paalanen}(1990)}]{Hebard1990}
\bibinfo{author}{\bibfnamefont{A.~F.} \bibnamefont{Hebard}} \bibnamefont{and}
  \bibinfo{author}{\bibfnamefont{M.~A.} \bibnamefont{Paalanen}},
  \bibinfo{journal}{Phys. Rev. Lett.} \textbf{\bibinfo{volume}{65}},
  \bibinfo{pages}{927} (\bibinfo{year}{1990}).

\bibitem[{\citenamefont{Goldman and Markovic}(1998)}]{Goldman1998}
\bibinfo{author}{\bibfnamefont{A.~M.} \bibnamefont{Goldman}} \bibnamefont{and}
  \bibinfo{author}{\bibfnamefont{N.}~\bibnamefont{Markovic}},
  \bibinfo{journal}{Physics Today} \textbf{\bibinfo{volume}{51}},
  \bibinfo{pages}{39} (\bibinfo{year}{1998}).

\bibitem[{\citenamefont{Orr et~al.}(1985)\citenamefont{Orr, Jaeger, and
  Goldman}}]{Orr1985}
\bibinfo{author}{\bibfnamefont{B.~G.} \bibnamefont{Orr}},
  \bibinfo{author}{\bibfnamefont{H.~M.} \bibnamefont{Jaeger}},
  \bibnamefont{and} \bibinfo{author}{\bibfnamefont{A.~M.}
  \bibnamefont{Goldman}}, \bibinfo{journal}{Phys. Rev. B.}
  \textbf{\bibinfo{volume}{32}}, \bibinfo{pages}{7586} (\bibinfo{year}{1985}).

\bibitem[{\citenamefont{Jaeger et~al.}(1986)\citenamefont{Jaeger, Haviland,
  Goldman, and Orr}}]{Jaeger1986}
\bibinfo{author}{\bibfnamefont{H.~M.} \bibnamefont{Jaeger}},
  \bibinfo{author}{\bibfnamefont{D.~B.} \bibnamefont{Haviland}},
  \bibinfo{author}{\bibfnamefont{A.~M.} \bibnamefont{Goldman}},
  \bibnamefont{and} \bibinfo{author}{\bibfnamefont{B.~G.} \bibnamefont{Orr}},
  \bibinfo{journal}{Phys. Rev. B.} \textbf{\bibinfo{volume}{34}},
  \bibinfo{pages}{4920} (\bibinfo{year}{1986}).

\bibitem[{\citenamefont{Okuma et~al.}(1998)\citenamefont{Okuma, Terashima, and
  Kokubo}}]{Okuma1998}
\bibinfo{author}{\bibfnamefont{S.}~\bibnamefont{Okuma}},
  \bibinfo{author}{\bibfnamefont{T.}~\bibnamefont{Terashima}},
  \bibnamefont{and} \bibinfo{author}{\bibfnamefont{N.}~\bibnamefont{Kokubo}},
  \bibinfo{journal}{Phys. Rev. B.} \textbf{\bibinfo{volume}{58}},
  \bibinfo{pages}{2816} (\bibinfo{year}{1998}).

\bibitem[{\citenamefont{Baturina et~al.}(2004)\citenamefont{Baturina, Islamov,
  Bentner, Strunk, Baklanov, and Satta}}]{Baturina2004}
\bibinfo{author}{\bibfnamefont{T.}~\bibnamefont{Baturina}},
  \bibinfo{author}{\bibfnamefont{D.}~\bibnamefont{Islamov}},
  \bibinfo{author}{\bibfnamefont{J.}~\bibnamefont{Bentner}},
  \bibinfo{author}{\bibfnamefont{C.}~\bibnamefont{Strunk}},
  \bibinfo{author}{\bibfnamefont{M.}~\bibnamefont{Baklanov}}, \bibnamefont{and}
  \bibinfo{author}{\bibfnamefont{A.}~\bibnamefont{Satta}},
  \bibinfo{journal}{JETP Lett.} \textbf{\bibinfo{volume}{79}},
  \bibinfo{pages}{337} (\bibinfo{year}{2004}).

\bibitem[{\citenamefont{Micolich et~al.}(2006)\citenamefont{Micolich, Tavenner,
  Powell, Hamilton, Curry, Giedd, and Meredith}}]{Micolich2006}
\bibinfo{author}{\bibfnamefont{A.~P.} \bibnamefont{Micolich}},
  \bibinfo{author}{\bibfnamefont{E.}~\bibnamefont{Tavenner}},
  \bibinfo{author}{\bibfnamefont{B.~J.} \bibnamefont{Powell}},
  \bibinfo{author}{\bibfnamefont{A.~R.} \bibnamefont{Hamilton}},
  \bibinfo{author}{\bibfnamefont{M.~T.} \bibnamefont{Curry}},
  \bibinfo{author}{\bibfnamefont{R.~E.} \bibnamefont{Giedd}}, \bibnamefont{and}
  \bibinfo{author}{\bibfnamefont{P.}~\bibnamefont{Meredith}},
  \bibinfo{journal}{Appl. Phys. Lett.} \textbf{\bibinfo{volume}{89}},
  \bibinfo{pages}{152503} (\bibinfo{year}{2006}).

\bibitem[{\citenamefont{Tavenner et~al.}(2004)\citenamefont{Tavenner, Meredith,
  Wood, Curry, and Giedd}}]{Tavenner2004}
\bibinfo{author}{\bibfnamefont{E.}~\bibnamefont{Tavenner}},
  \bibinfo{author}{\bibfnamefont{P.}~\bibnamefont{Meredith}},
  \bibinfo{author}{\bibfnamefont{B.}~\bibnamefont{Wood}},
  \bibinfo{author}{\bibfnamefont{M.}~\bibnamefont{Curry}}, \bibnamefont{and}
  \bibinfo{author}{\bibfnamefont{R.}~\bibnamefont{Giedd}},
  \bibinfo{journal}{Synth. Met.} \textbf{\bibinfo{volume}{145}},
  \bibinfo{pages}{183} (\bibinfo{year}{2004}).

\bibitem[{\citenamefont{Stephenson et~al.}(2009)\citenamefont{Stephenson,
  Divakar, Micolich, Meredith, and Powell}}]{Stephenson2009a}
\bibinfo{author}{\bibfnamefont{A.~P.} \bibnamefont{Stephenson}},
  \bibinfo{author}{\bibfnamefont{U.}~\bibnamefont{Divakar}},
  \bibinfo{author}{\bibfnamefont{A.~P.} \bibnamefont{Micolich}},
  \bibinfo{author}{\bibfnamefont{P.}~\bibnamefont{Meredith}}, \bibnamefont{and}
  \bibinfo{author}{\bibfnamefont{B.~J.} \bibnamefont{Powell}},
  \bibinfo{journal}{J. Appl. Phys.} \textbf{\bibinfo{volume}{105}},
  \bibinfo{pages}{093909} (\bibinfo{year}{2009}).

\bibitem[{\citenamefont{Strongin et~al.}(1970)\citenamefont{Strongin, Thompson,
  Kammerer, and Crow}}]{Strongin1970}
\bibinfo{author}{\bibfnamefont{M.}~\bibnamefont{Strongin}},
  \bibinfo{author}{\bibfnamefont{R.~S.} \bibnamefont{Thompson}},
  \bibinfo{author}{\bibfnamefont{O.~F.} \bibnamefont{Kammerer}},
  \bibnamefont{and} \bibinfo{author}{\bibfnamefont{J.~E.} \bibnamefont{Crow}},
  \bibinfo{journal}{Phys. Rev. B.} \textbf{\bibinfo{volume}{1}},
  \bibinfo{pages}{1078} (\bibinfo{year}{1970}).

\bibitem[{\citenamefont{Raffy et~al.}(1983)\citenamefont{Raffy, Laibowitz,
  Chaudari, and Maekawa}}]{Raffy1983}
\bibinfo{author}{\bibfnamefont{H.}~\bibnamefont{Raffy}},
  \bibinfo{author}{\bibfnamefont{R.~B.} \bibnamefont{Laibowitz}},
  \bibinfo{author}{\bibfnamefont{P.}~\bibnamefont{Chaudhari}}, \bibnamefont{and}
  \bibinfo{author}{\bibfnamefont{S.}~\bibnamefont{Maekawa}},
  \bibinfo{journal}{Phys. Rev. B.} \textbf{\bibinfo{volume}{28}},
  \bibinfo{pages}{6607} (\bibinfo{year}{1983}).

\bibitem[{\citenamefont{Graybeal and Beasley}(1984)}]{Graybeal1984}
\bibinfo{author}{\bibfnamefont{J.~M.} \bibnamefont{Graybeal}} \bibnamefont{and}
  \bibinfo{author}{\bibfnamefont{M.~R.} \bibnamefont{Beasley}},
  \bibinfo{journal}{Phys. Rev. B.} \textbf{\bibinfo{volume}{29}},
  \bibinfo{pages}{4167} (\bibinfo{year}{1984}).

\bibitem[{\citenamefont{Tinkham}(1963)}]{Tinkham1963}
\bibinfo{author}{\bibfnamefont{M.}~\bibnamefont{Tinkham}},
  \bibinfo{journal}{Phys. Rev.} \textbf{\bibinfo{volume}{129}},
  \bibinfo{pages}{2413} (\bibinfo{year}{1963}).

\bibitem[{\citenamefont{Kobayashi et~al.}(1981)\citenamefont{Kobayashi, Tada,
  and Sasaki}}]{Kobayashi1981}
\bibinfo{author}{\bibfnamefont{S.}~\bibnamefont{Kobayashi}},
  \bibinfo{author}{\bibfnamefont{Y.}~\bibnamefont{Tada}}, \bibnamefont{and}
  \bibinfo{author}{\bibfnamefont{W.}~\bibnamefont{Sasaki}},
  \bibinfo{journal}{Physica B} \textbf{\bibinfo{volume}{107}},
  \bibinfo{pages}{129} (\bibinfo{year}{1981}).

\bibitem[{\citenamefont{White et~al.}(1986)\citenamefont{White, Dynes, and
  Garno}}]{White1986}
\bibinfo{author}{\bibfnamefont{A.~E.} \bibnamefont{White}},
  \bibinfo{author}{\bibfnamefont{R.~C.} \bibnamefont{Dynes}}, \bibnamefont{and}
  \bibinfo{author}{\bibfnamefont{J.~P.} \bibnamefont{Garno}},
  \bibinfo{journal}{Phys. Rev. B} \textbf{\bibinfo{volume}{33}},
  \bibinfo{pages}{3549} (\bibinfo{year}{1986}).

\bibitem[{\citenamefont{Kunchur et~al.}(1987)\citenamefont{Kunchur, Zhang,
  Lindenfeld, McLean, and Brooks}}]{Kunchur1987}
\bibinfo{author}{\bibfnamefont{M.}~\bibnamefont{Kunchur}},
  \bibinfo{author}{\bibfnamefont{Y.~Z.} \bibnamefont{Zhang}},
  \bibinfo{author}{\bibfnamefont{P.}~\bibnamefont{Lindenfeld}},
  \bibinfo{author}{\bibfnamefont{W.~L.} \bibnamefont{McLean}},
  \bibnamefont{and} \bibinfo{author}{\bibfnamefont{J.~S.}
  \bibnamefont{Brooks}}, \bibinfo{journal}{Phys. Rev. B.}
  \textbf{\bibinfo{volume}{36}}, \bibinfo{pages}{4062} (\bibinfo{year}{1987}).

\bibitem[{\citenamefont{Jaeger et~al.}(1989)\citenamefont{Jaeger, Haviland,
  Orr, and Goldman}}]{Jaeger1989}
\bibinfo{author}{\bibfnamefont{H.~M.} \bibnamefont{Jaeger}},
  \bibinfo{author}{\bibfnamefont{D.~B.} \bibnamefont{Haviland}},
  \bibinfo{author}{\bibfnamefont{B.~G.} \bibnamefont{Orr}}, \bibnamefont{and}
  \bibinfo{author}{\bibfnamefont{A.~M.} \bibnamefont{Goldman}},
  \bibinfo{journal}{Phys. Rev. B} \textbf{\bibinfo{volume}{40}},
  \bibinfo{pages}{182} (\bibinfo{year}{1989}).

\bibitem[{\citenamefont{Gerber et~al.}(1990)\citenamefont{Gerber, Grenet,
  Cyrot, and Beille}}]{Gerber1990}
\bibinfo{author}{\bibfnamefont{A.}~\bibnamefont{Gerber}},
  \bibinfo{author}{\bibfnamefont{T.}~\bibnamefont{Grenet}},
  \bibinfo{author}{\bibfnamefont{M.}~\bibnamefont{Cyrot}}, \bibnamefont{and}
  \bibinfo{author}{\bibfnamefont{J.}~\bibnamefont{Beille}},
  \bibinfo{journal}{Phys. Rev. Lett.} \textbf{\bibinfo{volume}{65}},
  \bibinfo{pages}{3201} (\bibinfo{year}{1990}).

\bibitem[{\citenamefont{Gray and Kim}(1993)}]{Gray1993}
\bibinfo{author}{\bibfnamefont{K.~E.} \bibnamefont{Gray}} \bibnamefont{and}
  \bibinfo{author}{\bibfnamefont{D.~H.} \bibnamefont{Kim}},
  \bibinfo{journal}{Phys. Rev. Lett.} \textbf{\bibinfo{volume}{70}},
  \bibinfo{pages}{1693} (\bibinfo{year}{1993}).

\bibitem[{\citenamefont{Kartsovnik et~al.}(1999)\citenamefont{Kartsovnik,
  Logvenov, and Kushch}}]{Kartsovnik1999}
\bibinfo{author}{\bibfnamefont{M.~V.} \bibnamefont{Kartsovnik}},
  \bibinfo{author}{\bibfnamefont{G.~Y.} \bibnamefont{Logvenov}},
  \bibnamefont{and} \bibinfo{author}{\bibfnamefont{N.~D.}
  \bibnamefont{Kushch}}, \bibinfo{journal}{Synth. Met.}
  \textbf{\bibinfo{volume}{103}}, \bibinfo{pages}{1827} (\bibinfo{year}{1999}).

\bibitem[{\citenamefont{Bergmann}(1984)}]{Bergmann1984}
\bibinfo{author}{\bibfnamefont{G.}~\bibnamefont{Bergmann}},
  \bibinfo{journal}{Phys. Rep.} \textbf{\bibinfo{volume}{107}},
  \bibinfo{pages}{1} (\bibinfo{year}{1984}).

\bibitem[{\citenamefont{Osaheni et~al.}(1992)\citenamefont{Osaheni, Jenekhe,
  Burns, Du, Joo, Epstein, and Wang}}]{Osaheni1992}
\bibinfo{author}{\bibfnamefont{J.~A.} \bibnamefont{Osaheni}},
  \bibinfo{author}{\bibfnamefont{S.~A.} \bibnamefont{Jenekhe}},
  \bibinfo{author}{\bibfnamefont{A.}~\bibnamefont{Burns}},
  \bibinfo{author}{\bibfnamefont{G.}~\bibnamefont{Du}},
  \bibinfo{author}{\bibfnamefont{J.}~\bibnamefont{Joo}},
  \bibinfo{author}{\bibfnamefont{A.~J.} \bibnamefont{Epstein}},
  \bibnamefont{and} \bibinfo{author}{\bibfnamefont{C.~S.} \bibnamefont{Wang}},
  \bibinfo{journal}{Macromolecules} \textbf{\bibinfo{volume}{25}},
  \bibinfo{pages}{5828} (\bibinfo{year}{1992}).

\bibitem[{\citenamefont{Zuo et~al.}(1996)\citenamefont{Zuo, Schlueter, Kelly,
  and Williams}}]{Zuo1996}
\bibinfo{author}{\bibfnamefont{F.}~\bibnamefont{Zuo}},
  \bibinfo{author}{\bibfnamefont{J.~A.} \bibnamefont{Schlueter}},
  \bibinfo{author}{\bibfnamefont{M.~E.} \bibnamefont{Kelly}}, \bibnamefont{and}
  \bibinfo{author}{\bibfnamefont{J.~M.} \bibnamefont{Williams}},
  \bibinfo{journal}{Phys. Rev. B} \textbf{\bibinfo{volume}{54}},
  \bibinfo{pages}{11 973} (\bibinfo{year}{1996}).

\bibitem[{\citenamefont{Zuo et~al.}(1999)\citenamefont{Zuo, Schlueter, and
  Williams}}]{Zuo1999}
\bibinfo{author}{\bibfnamefont{F.}~\bibnamefont{Zuo}},
  \bibinfo{author}{\bibfnamefont{J.~A.} \bibnamefont{Schlueter}},
  \bibnamefont{and} \bibinfo{author}{\bibfnamefont{J.~M.}
  \bibnamefont{Williams}}, \bibinfo{journal}{Phys. Rev. B}
  \textbf{\bibinfo{volume}{60}}, \bibinfo{pages}{574} (\bibinfo{year}{1999}).

\bibitem[{\citenamefont{Singleton}(2000)}]{Singleton2000}
\bibinfo{author}{\bibfnamefont{J.}~\bibnamefont{Singleton}},
  \bibinfo{journal}{Rep. Prog. Phys.} \textbf{\bibinfo{volume}{63}},
  \bibinfo{pages}{1111} (\bibinfo{year}{2000}).

\bibitem[{\citenamefont{Analytis et~al.}(2006)\citenamefont{Analytis, Ardavan,
  Blundell, Owen, Garman, Jeynes, and Powell}}]{Analytis2006}
\bibinfo{author}{\bibfnamefont{J.~G.} \bibnamefont{Analytis}},
  \bibinfo{author}{\bibfnamefont{A.}~\bibnamefont{Ardavan}},
  \bibinfo{author}{\bibfnamefont{S.~J.} \bibnamefont{Blundell}},
  \bibinfo{author}{\bibfnamefont{R.~L.} \bibnamefont{Owen}},
  \bibinfo{author}{\bibfnamefont{E.~F.} \bibnamefont{Garman}},
  \bibinfo{author}{\bibfnamefont{C.}~\bibnamefont{Jeynes}}, \bibnamefont{and}
  \bibinfo{author}{\bibfnamefont{B.~J.} \bibnamefont{Powell}},
  \bibinfo{journal}{Phys. Rev. Lett.} \textbf{\bibinfo{volume}{96}},
  \bibinfo{pages}{177002} (\bibinfo{year}{2006}).

\bibitem[{\citenamefont{Powell and McKenzie}(2004)}]{Powell2004}
\bibinfo{author}{\bibfnamefont{B.~J.} \bibnamefont{Powell}} \bibnamefont{and}
  \bibinfo{author}{\bibfnamefont{R.~H.} \bibnamefont{McKenzie}},
  \bibinfo{journal}{Phys. Rev. B} \textbf{\bibinfo{volume}{69}},
  \bibinfo{pages}{024519} (\bibinfo{year}{2004}).

\bibitem[{\citenamefont{Powell and McKenzie}(2006)}]{Powell2006}
\bibinfo{author}{\bibfnamefont{B.~J.} \bibnamefont{Powell}} \bibnamefont{and}
  \bibinfo{author}{\bibfnamefont{R.~H.} \bibnamefont{McKenzie}},
  \bibinfo{journal}{J. Phys.: Condens. Matter} \textbf{\bibinfo{volume}{18}},
  \bibinfo{pages}{R827} (\bibinfo{year}{2006}).

\end{thebibliography}

\end{document}